\documentclass[a4paper,11pt]{article}
\usepackage{bm,amsmath,amssymb,mathrsfs,graphicx,citesort}
\newcommand{\nc}{\newcommand}
\nc{\rnc}{\renewcommand}
\nc{\beq}{\begin{equation}}
\nc{\eeq}{\end{equation}}
\nc{\nn}{\nonumber}
\rnc{\Im}{{\rm{Im}\,}}
\rnc{\Re}{{\rm{Re}\,}}
\rnc{\i}{{\rm i}}
\rnc{\d}{{\rm d}}
\nc{\e}{{\rm e}}
\nc{\mfa}{{\mathfrak{a}}}
\nc{\mfab}{\overline{\mfa}}
\nc{\mfA}{{\mathfrak{A}}}
\nc{\mfAb}{\overline{\mfA}}
\nc{\mfb}{{\mathfrak{b}}}
\nc{\mfbb}{\overline{\mfb}}
\nc{\mfB}{{\mathfrak{B}}}
\nc{\mfBb}{\overline{\mfB}}
\nc{\mfc}{{\mathfrak{c}}}
\nc{\mfcb}{\overline{\mfc}}
\nc{\mfC}{{\mathfrak{C}}}
\nc{\mfCb}{\overline{\mfC}}
\nc{\db}{\displaybreak[0]\\}
\nc{\bra}{\langle}
\nc{\ket}{\rangle}
\nc{\bbra}{\left\langle}
\nc{\kket}{\right\rangle}
\nc{\J}{\mathscr{J}}
\nc{\R}{\mathscr{R}}
\nc{\T}{\mathscr{T}}
\rnc{\H}{\mathscr{H}}
\nc{\M}{\mathscr{M}}
\nc{\Q}{\mathscr{Q}}
\nc{\A}{\mathscr{A}}
\nc{\B}{\mathscr{B}}
\nc{\C}{\mathscr{C}}
\nc{\D}{\mathscr{D}}
\nc{\Z}{\mathcal{Z}}
\nc{\ep}{\varepsilon}
\nc{\chem}{\mu_{\rm c}}
\rnc{\(}{\left(}
\rnc{\)}{\right)}
\rnc{\[}{\left[}
\rnc{\]}{\right]}
\nc{\dprime}{\prime\prime}
\nc{\bi}{{\bar{i}}}
\nc{\tmfb}{\widetilde{\mfb}}
\nc{\bmfb}{\widehat{\mfb}}
\nc{\bGamma}{\widehat{\Gamma}}
\nc{\tGamma}{\widetilde{\Gamma}}
\nc{\XXZ}{$XXZ$\,}
\nc{\sotimes}{\mathop{\otimes}_{\rm s}}

\DeclareMathOperator{\sh}{sh}
\DeclareMathOperator{\ch}{ch}

\newtheorem{theorem}{Theorem}
\newtheorem{proposition}{Proposition}

\newtheorem{corollary}{Corollary}

\numberwithin{equation}{section}
\numberwithin{lemma}{section}
\numberwithin{proposition}{section}
\numberwithin{theorem}{section}
\numberwithin{corollary}{section}

\begin{document}%
%
\title{Form factors and correlation functions of \\an interacting spinless fermion model}
\author{Kohei Motegi\thanks{E-mail: motegi@gokutan.c.u-tokyo.ac.jp} \,
and  Kazumitsu Sakai\thanks{E-mail: sakai@gokutan.c.u-tokyo.ac.jp}\\\\
\it Institute of physics, University of Tokyo, \\
\it Komaba 3-8-1,
Meguro-ku, Tokyo 153-8902, Japan \\\\
\\}

\date{August 8, 2007}
 
 
\maketitle

%
%
\begin{abstract}
Introducing the fermionic $R$-operator and solutions of the 
inverse scattering problem for local fermion operators, 
we derive a multiple integral representation for  zero-temperature
correlation functions of a one-dimensional interacting spinless 
fermion model.
Correlation functions  particularly considered are the  
one-particle Green's function and the density-density correlation 
function both for
any interaction strength and for arbitrary particle densities.
In particular for the free fermion model,
our formulae reproduce the known exact results. 
Form factors of local fermion operators are also 
calculated for a finite system.
\end{abstract}

%
%
\section{Introduction}

The evaluation of correlation functions has been one 
of the challenging problems in research on  quantum 
many-body systems in one-dimension, since most of the 
intriguing phenomena induced by underlying strong quantum 
fluctuations are theoretically described through  
correlation functions. 
In these systems, it is widely known that the existence 
of models which are solvable by means of  Bethe ansatz
(see \cite{KBIbook,Takabook} for example). 
Though the exact computation of  correlation functions, 
of course, is still tremendously difficult even in such 
models, several analytical methods have been recently  
developed to derive manageable expressions for correlation 
functions, especially in the spin-1/2 Heisenberg XXZ chain.

In 1990s, correlation functions of the spin-1/2 XXZ 
chain at zero temperature and for zero magnetic field have 
been expressed as  multiple integral forms derived by 
the $q$-vertex operator approach \cite{JMMN,JMbook,JM}.
An alternative method combining the algebraic
Bethe ansatz with  solutions to the quantum inverse
problem  has been provided
for the XXZ chain in arbitrary magnetic fields
\cite{KMT00,KMST02,KMST05-1} (see also \cite{KMSTreview} for 
a recent review). 
This method can be generalized to the finite-temperature 
and/or the time-dependent correlation functions
\cite{KMST05,GKS04,GKS05,GHS05,Sakai07}.

In general, one-dimensional (1D) quantum spin systems
are mapped to 1D fermion systems through the Jordan-Wigner 
transformation. For the spin-1/2 XXZ chain, the 
corresponding system is a spinless fermion 
model with the nearest neighbor hopping and interaction.
In the thermodynamic limit, the bulk quantities in the 
XXZ chain are exactly the same as those in the spinless 
fermion model. Namely, in the 1D quantum systems, the 
difference of the statistics between the spin and the
fermion does not show up,  as long as we 
concentrate on their bulk quantities.

The situation, however, is radically changed when the
quantities accompanying a change of the number of particles
are considered.
For instance, let us consider the (equal-time) one-particle
Green's function $\bra c_1 c_{m+1}^{\dagger}\ket$ for the spinless 
fermion model  and the transverse spin-spin 
correlation function $\bra \sigma_1^+\sigma_{m+1}^-\ket$ for 
the XXZ chain, which are intuitively the same. 
In fact, due to the difference of the statistics, or 
equivalently the nonlocality
of the Jordan-Wigner transformation, both the 
correlation functions exhibit completely different 
behavior. 
For instance, one observes an oscillatory behavior
for the one-particle Green's function (referred to 
as the $k_{\rm F}$-oscillation, where $k_{\rm F}$ 
is the Fermi momentum),  which is peculiar to the 
fermion systems. In contrast, for the transverse 
spin-spin correlation function, such an oscillatory 
behavior does not appear\footnote{
In fact, there could exist an oscillatory behavior
in the transverse spin-spin correlation function,
which, however, can be eliminated by a gauge transformation.}.

As mentioned above,  exact expressions for the correlation 
functions of the XXZ chain have already
been  proposed in the form of  multiple integral 
representations. 
Unfortunately, once the Jordan-Wigner transformation 
is performed and the XXZ chain is considered instead
of the spinless fermion model, 
it is difficult to trace the difference 
of the statistics in the framework of 
multiple integral representations. 
Namely, to derive a manageable expression of 
correlation functions for the spinless fermion model,
we must directly treat the fermion system from the 
very beginning.

 In this paper,  introducing the fermionic $R$-operator 
\cite{USW} which acts on the fermion Fock space, we directly treat 
the spinless fermion model without mapping to the XXZ chain. 
Combining the method provided in the XXZ chain \cite{KMST02} 
with  solutions to the inverse scattering problem 
of local fermion operators \cite{GK00}, we derive a multiple 
integral representing the equal-time one-particle Green's function 
and the density-density correlation function at zero temperature 
both for any interaction strengths and for arbitrary particle 
densities.
Our formulae reproduce the known results
for the free fermion model.
In addition to the correlation functions, we also compute
form factors for local fermion operators, which might be
useful for systematic evaluations of the spectral 
functions for the spinless fermion model.

This paper is organized as follows. In the subsequent section, 
we introduce the fermionic $R$-operator and the 
transfer operator constructed by the $R$-operator.
Then we briefly review the algebraic Bethe ansatz for 
the spinless fermion model. 
The scalar product of a Bethe state with an arbitrary 
state is presented in section 3. 
Combining  solutions of the inverse scattering
problem with the scalar product, we compute form factors of
local fermion operators.
Multiple integral representations for correlation 
functions are derived in section 4.  
In section 5, using the multiple integral representations,
we explicitly evaluate correlation functions for the
free fermion model.
Section 6 is devoted to a brief conclusion.
%
%
\section{Spinless fermion model}
%
The Hamiltonian of the spinless fermion model on a periodic lattice with $M$ 
sites is defined as
\begin{align}
&H=H_{0}-\chem \sum_{j=1}^{M}\(\frac{1}{2}-n_{j}\), \nonumber \\
&H_{0}=t\sum_{j=1}^{M}\left\{ c_{j}^{\dagger}c_{j+1}+c_{j+1}^{\dagger}
                              c_{j}+2\Delta\(\(\frac{1}{2}-n_{j}\)
                                  \(\frac{1}{2}-n_{j+1}\)-\frac{1}{4}\)
                     \right\},
\label{Hamiltonian}
\end{align}
where $c_{j}^{\dagger}$ and $c_{j}$ are the fermionic creation and 
annihilation operators at the $j$th site, respectively, satisfying the 
canonical anti-commutation relations
\begin{eqnarray}
\{ c_{j},c_{k} \}=\{ c_{j}^{\dagger},c_{k}^{\dagger} \}=0, \ \ \{ c_{j}^{\dagger}, c_{k}\}=\delta_{jk}.
\end{eqnarray}
Here $t$ and $\Delta$ are real  constants characterizing
the nature of the ground state, and $\chem$ denotes the
chemical potential coupling to the density operator 
$n_j=c_j^{\dagger}c_j$.
By use of the Jordan-Wigner transformation
\begin{equation}
c_j=\exp\[\i\pi\sum_{k=1}^{j-1}n_k\]\sigma_j^+,\quad
c_j^{\dagger}=\exp\[-\i\pi\sum_{k=1}^{j-1}n_k\]\sigma_j^-, \quad
n_j=\frac{1}{2}(1-\sigma_j^z),
\label{JW}
\end{equation}
the spinless fermion model \eqref{Hamiltonian} can be mapped 
to the spin-1/2 XXZ chain with an external magnetic field $h$:
\begin{equation}
H_{\rm XXZ}=J\sum_{j=1}^M
    \left\{\sigma_j^+\sigma_{j+1}^-+
     \sigma_{j+1}^+\sigma_j^-+
    \frac{\Delta}{2}(\sigma_j^z\sigma_{j+1}^z-1)\right\}
 -\frac{h}{2}\sum_{j=1}^M\sigma_j^z,
\label{xxz}
\end{equation}
where we have changed the variables $t$ and $\mu_{\rm c}$ 
in \eqref{Hamiltonian} as $t\to J$ and $\chem\to h$.

The difference between the spinless fermion model
and the corresponding XXZ chain lies only in the
difference of boundary conditions. 
Therefore, in the thermodynamic limit, the quantities 
without accompanying a change of the number of particles 
such as the density-density correlation function 
$\bra n_1 n_{m+1} \ket$ are exactly the same as
those such as a longitudinal spin-spin correlation 
function $\bra(1-\sigma_1^z)(1-\sigma_{m+1}^z) \ket$/4 for the 
XXZ chain. 
As mentioned in the preceding section, however, the 
quantities changing the number of particles such as 
the one-particle Green's function $\bra c_1 c_{m+1}^{\dagger}  \ket$ 
exhibit completely different behavior 
from those such as the transverse spin-spin 
correlation function $\bra \sigma_1^+\sigma_{m+1}^-\ket$,
due to the nonlocality of the Jordan-Wigner 
transformation \eqref{JW}.
Though the exact expression of $\bra\sigma_1^+\sigma_{m+1}^- \ket$
for the XXZ chain has already been presented in \cite{KMST02},
it is difficult to convert it to the expression for  the spinless
fermion model, as long as we concentrate on the XXZ chain \eqref{xxz}. 
Namely we should treat the original spinless 
fermion model without mapping to the XXZ chain. 
In this section, 
introducing the fermionic $R$-operator \cite{USW}, we directly 
consider the spinless fermion model \eqref{Hamiltonian}.

The fermionic $R$-operator is defined by 
\begin{equation}
R_{12}(\lambda)=1-n_{1}-n_{2}+\frac{\sh\lambda}{\sh(\lambda+\eta)}(n_1+n_2-2 n_1 n_2)
 +\frac{\sh\eta}{\sh(\lambda+\eta)}
 (c_{1}^{\dagger}c_{2}+c_{2}^{\dagger}c_{1}),
\label{fermion}
\end{equation}
which acts on $V_{1} \sotimes V_{2}$. Here $V_{j}$ is a 
two-dimensional fermion Fock space whose normalized orthogonal
basis is given by
$|0 \ket _{j}$ and $|1 \ket _{j} := c_{j}^{\dagger} | 0\ket_{j}$,
where $ c_{j}|0 \ket_{j}=0$  and  $\sotimes$ denotes the super tensor product.
Note that this fermionic $R$-operator satisfies the Yang-Baxter equation
\cite{USW}:
\begin{equation}
R_{12}(\lambda_{1}-\lambda_{2})R_{13}(\lambda_{1})R_{23}(\lambda_{2})
=R_{23}(\lambda_{2})R_{13}(\lambda_{1})R_{12}(\lambda_{1}-\lambda_{2}).
\label{YBE}
\end{equation}
Identifying one of the two fermionic Fock spaces with the quantum 
space $\mathcal{H}_m$, we define the $L$-operator at the $m$th site by
\begin{equation}
L_{m}(\lambda)=R_{am}(\lambda-\xi_m),
\label{L-op}
\end{equation}
where $\xi_m$ are inhomogeneous parameters assumed to 
be arbitrary complex numbers.
The fermionic monodromy operator is then constructed as 
a product of the $L$-operators:
\begin{equation}
T(\lambda)=L_{M}(\lambda)\cdots L_{2}(\lambda)L_{1}(\lambda)=
A(\lambda)(1-n_{a})+B(\lambda)c_{a}+c_{a}^{\dagger}C(\lambda)+D(\lambda)n_{a}.
\label{monodromy}
\end{equation}
Thanks to the Yang-Baxter equation \eqref{YBE}, the fermionic 
transfer operator defined by
\begin{equation}
\mathcal{T}(\lambda)=\text{str}_{a}T(\lambda) =
{}_{a} \bra 0|T(\lambda) |0 \ket_{a} -{}_{a}\bra 1|T(\lambda) |1 \ket_{a}
=A(\lambda)-D(\lambda),
\label{TM}
\end{equation}
constitutes a commuting family:
$[\mathcal{T}(\lambda), \mathcal{T}(\mu)]=0$,
where the dual fermion Fock space is spanned 
by $_{a}\bra 0|$ and $_{a}\bra 1|$ with
${}_{a}\bra 1|:= {}_{a} \bra 0|c_{a}$
and
${}_{a}\bra 0|c_{a}^{\dagger}=0$.

The Hamiltonian of the spinless fermion model $H_0$ 
\eqref{Hamiltonian} can be expressed as the logarithmic
derivative of the fermionic transfer operator $\mathcal{T}(\lambda)$
in the homogeneous limit $\xi_m\to\eta/2$:
\begin{equation}
H_{0}=t \sh (\eta) \frac{\partial}{\partial \lambda}
        \ln \mathcal{T}(\lambda)\biggr|_{\lambda=\xi_m=\frac{\eta}{2}},
\quad \Delta=\ch \eta.
\label{Baxter}
\end{equation}
Due to the Yang-Baxter equation \eqref{YBE}, one immediately
sees that the following  relation is valid for arbitrary spectral 
parameters $\lambda_1$ and $\lambda_2$:
\begin{equation}
R_{12}(\lambda_{1}-\lambda_{2})T_{1}(\lambda_{1})T_{2}(\lambda_{2})
=T_{2}(\lambda_{2})T_{1}(\lambda_{1})R_{12}(\lambda_{1}-\lambda_{2}).
\end{equation}
This leads to the commutation relations among the operators
$A(\lambda)$, $B(\lambda)$, $C(\lambda)$ and $D(\lambda)$ 
constructing the
monodromy operator \eqref{monodromy}. 
Let us explicitly write down  some of these relations, which we will 
use later:
\begin{align}
&B(\mu)B(\lambda)=B(\lambda)B(\mu), \ \ \ C(\mu)C(\lambda)=C(\lambda)C(\mu), 
                                                                \nn \\
&A(\mu)B(\lambda)=f(\lambda, \mu)B(\lambda)A(\mu)-g(\lambda,\mu)B(\mu)A(\lambda), 
                                                                \nn \\
&D(\mu)B(\lambda)=-f(\mu,\lambda)B(\lambda)D(\mu)+g(\mu,\lambda)B(\mu)D(\lambda), 
                                                                \nn \\
&C(\mu)A(\lambda)=f(\mu,\lambda)A(\lambda)C(\mu)-g(\mu,\lambda)A(\mu)C(\lambda), 
                                                                \nn \\
&C(\mu)D(\lambda)=-f(\lambda,\mu)D(\lambda)C(\mu)+g(\lambda,\mu)D(\mu)C(\lambda), 
                                                                \nn \\
&C(\mu)B(\lambda)+B(\lambda)C(\mu)=g(\lambda,\mu)(D(\lambda)A(\mu)-D(\mu)A(\lambda)),
\label{commutation}
\end{align}
where
\begin{equation}
f(\lambda, \mu)=\frac{\sh(\lambda-\mu+\eta)}{\sh(\lambda-\mu)}, 
\quad  g(\lambda, \mu)=\frac{\sh\eta}{\sh(\lambda-\mu)}.
\end{equation}

Let $|\Omega\ket$ be the vacuum state for the system \eqref{Hamiltonian}.
Obviously $|\Omega\ket$ is an eigenstate of the transfer operator
$\mathcal{T}(\mu)$:
\begin{equation}
\mathcal{T}(\mu)|\Omega\ket=(a(\mu)-d(\mu))|\Omega\ket,
\quad a(\mu)=1,\quad d(\mu)=\prod_{m=1}^M f^{-1}(\mu,\xi_m).
\end{equation}
A general $N$-particle state $|\psi\ket$ can be constructed by 
a multiple action of  $B(\lambda)$ on $|\Omega\ket$ as
\begin{equation}
| \psi \ket =\prod_{j=1}^{N}B(\lambda_{j})|\Omega \ket.  
 \quad  N=0, 1, \dots, M.
\label{state}
\end{equation}
Utilizing  commutation relations in \eqref{commutation},
one finds that the state \eqref{state} becomes 
an eigenstate of $\mathcal{T}(\mu)$, if the complex 
parameters $\{\lambda_j \}$ satisfy the Bethe 
ansatz equation;
\begin{equation}
(-1)^{N+1}=\frac{d(\lambda_j)}{a(\lambda_j)}
  \prod_{\substack{k=1 \\k \neq j}}^{N}
            \frac{f(\lambda_j,\lambda_k)}
                 {f(\lambda_k,\lambda_j)}=
\prod_{m=1}^{M}\frac{\sh(\lambda_{j}-\xi_{m})}{\sh(\lambda_{j}-\xi_{m}+\eta)}
 \prod_{\substack{k=1 \\k \neq j}}^{N}\frac{\sh(\lambda_{j}-\lambda_{k}+\eta)}
                               {\sh(\lambda_{j}-\lambda_{k}-\eta)}.
\label{BAE}
\end{equation}
Then the corresponding eigenvalue $\tau(\mu)$ (i.e. 
$\mathcal{T}(\mu)|\psi\ket=\tau(\mu) |\psi\ket$)
is given by
\begin{equation}
\tau(\mu)=a(\mu)\prod_{j=1}^{N}f(\lambda_{j},\mu)+
                        (-1)^{N+1}d(\mu)\prod_{j=1}^{N} f(\mu,\lambda_{j}).
\label{eigen}
\end{equation}
Note that the existence of an extra factor $(-1)^{N+1}$  
in \eqref{BAE} and \eqref{eigen} reflects the fermionic structure 
of the system, which does not appear for the XXZ chain \eqref{xxz}.
The relation \eqref{Baxter} leads to the explicit expression of the
energy spectrum per site $e_0$ of the spinless fermion model:
\begin{equation}
e_0=\frac{1}{M}\sum_{j=1}^N \frac{t \sh^2\eta}{\sh(\lambda_j+\frac{\eta}{2})
               \sh(\lambda_j-\frac{\eta}{2})}-\chem\(\frac{1}{2}-\frac{N}{M}\).
\label{energy}
\end{equation}

In the thermodynamic limit, where $M \to \infty $, $N\to \infty$, and $N/M$ 
is a constant determined by a function of the chemical potential $\chem$,
the Bethe ansatz equation \eqref{BAE} characterizing the ground state
reduces to the following linear integral equation for the distribution
function of the Bethe roots:

\begin{equation}
-2\pi \i \rho_{\rm tot}(\lambda)+
\int_{\mathcal{C}}K(\lambda-\mu)\rho_{\rm tot}(\mu)\d\mu=t\(\lambda,\frac{\eta}{2}\).
\label{density}
\end{equation}
In the above, the integral kernel $K(\lambda)$ and the 
function $t(\lambda,\mu)$ are respectively defined by
\begin{eqnarray}
K(\lambda)=\frac{\sh 2\eta}{\sh(\lambda+\eta)\sh(\lambda-\eta)}, \quad
t(\lambda,\mu)=\frac{\sh\eta}{\sh(\lambda-\mu)\sh(\lambda-\mu+\eta)}.
\end{eqnarray}
The integration contour $\mathcal{C}=[-\Lambda_{\chem}, \Lambda_{\chem}]$ is
given by $\ep(\pm \Lambda_{\chem})=0$, where $\ep(\lambda)$
is the dressed energy satisfying the following linear integral
equation:
\begin{equation}
-2\pi\i\ep(\lambda)+\int_{\mathcal{C}} K(\lambda-\mu)\ep(\mu)\d \mu
                      =t\sh(\eta) t\(\lambda,\frac{\eta}{2}\)+\chem.
\label{dress}
\end{equation}
Namely, for $-1<\Delta \le 1$ ($\eta=-\i \zeta$, $\zeta>0$), 
the contour $\mathcal{C}$ is defined by an interval on the real axis.
In particular, at $\chem\to 0$, $\Lambda_{\chem}\to \infty$.
On the other hand, for $\Delta>1$ $(\eta<0)$, the Bethe 
roots are distributed on the imaginary axis. 
In particular, at $\chem=0$, $\Lambda_{\chem}=-\pi \i/2$.
Then the energy density \eqref{energy} explicitly reads
\begin{equation}
e_0=t\sh(\eta)\int_{\mathcal{C}} 
                    t\(\lambda,\frac{\eta}{2}\)\rho_{\rm tot}(\lambda)\d \lambda
-\chem\(\frac{1}{2}-\bra n_j\ket\)
 =\int_{\mathcal{C}} t\(\lambda,\frac{\eta}{2}\) \ep(\lambda)\d \lambda 
                                -\frac{\chem}{2}, 
\label{energy2}
\end{equation}
where $\bra n_j \ket$ is the particle density given by
\begin{equation}
\bra n_j \ket=\int_{\mathcal{C}}\rho_{\rm tot}(\lambda)\d \lambda.
\label{particle-density}
\end{equation}
Note that the above expression of the ground state energy \eqref{energy2},
the distribution function \eqref{density} and the dressed energy
\eqref{dress} are exactly the same as those for the XXZ chain 
\cite{Takabook,KBIbook}.

For later convenience, let us define the inhomogeneous distribution
function $\rho(\lambda,\xi)$ \cite{KMST02} as the solution of the
integral equation:
\begin{equation}
-2\pi \i \rho(\lambda,\xi)+\int_{\mathcal{C}}K(\lambda-\mu)
             \rho(\mu,\xi)\d\mu =t(\lambda,\xi).
\label{inhom-density}
\end{equation}
Correspondingly, the inhomogeneous total distribution function 
$\rho_{\rm tot}(\lambda,\{\xi\})$ is defined by
\begin{equation}
\rho_{\rm tot}(\lambda,\{\xi\})=\frac{1}{M}\sum_{m=1}^M \rho(\lambda,\xi_m),
\label{total-density}
\end{equation}
which reduces to $\rho_{\rm tot}(\lambda)$ 
\eqref{density}, when one takes the homogeneous limit 
$\xi_m\to \eta/2$.

In particular, at zero chemical potential $\mu_{\rm c}=0$
(corresponding to the half filling case 
$\bra n_j \ket=1/2$), one has
\begin{align}
\rho(\lambda,\xi)=
  \begin{cases}
    \frac{1}{2\zeta \ch\frac{\pi}{\zeta}(\lambda+\frac{\eta}{2}-\xi)}& 
                           \text{for $|\Delta | <1, \zeta =\i$} \eta \\
    \frac{\i}{2\pi}{\sum_{n=-\infty}^{\infty}}
                           \frac{\e^{2n(\lambda+\frac{\eta}{2}-\xi)}}{\ch(n\eta)}=
    \frac{\i}{2\pi}{\prod_{n=1}^{\infty}} 
                   \( \frac{1-q^{2n}}{1+q^{2n}}\)^{2}
                    \frac{\vartheta_{3}(i(\lambda+\frac{\eta}{2}-\xi),q)}
                          {\vartheta_{4}(i(\lambda+\frac{\eta}{2}-\xi),q)}& 
                           \text{for $\Delta >1, q=e^{\eta}$}
  \end{cases},
\label{distribution}
\end{align}
where $\vartheta_n(\lambda,q)$ is the Jacobi theta function.
%
\section{Scalar products and form factors}
%
\subsection{Scalar Products}
%
Our main aim is to evaluate the correlation 
function of the spinless fermion model \eqref{Hamiltonian}:
\begin{equation}
\bra \mathcal{O}_1^{\dagger}\mathcal{O}_{m+1}\ket
=\frac{\bra \psi_{\rm g}|\mathcal{O}_1^{\dagger} 
                              \mathcal{O}_{m+1}|\psi_{\rm g}\ket}
      {\bra \psi_{\rm g}|\psi_{\rm g} \ket},
\label{correlation}
\end{equation}
where $|\psi_{\rm g} \ket$ is the ground state constructed by substituting
an appropriate solution  $\{\lambda\}$ of the Bethe ansatz equation \eqref{BAE} 
into  \eqref{state}, and $\bra \psi_{\rm g}|$ is the dual state:
$\bra \psi_{\rm g}|=\bra \Omega|\prod_{j=1}^{N} C(\lambda_j)$. 
In this paper we concentrate on the one-particle Green's function 
($\mathcal{O}_j=c_j, c_j^{\dagger}$) and the density-density correlation function 
($\mathcal{O}_j=n_j$). Here and in what follows, we consider the 
``inhomogeneous" model \eqref{monodromy} for convenience. The correlation functions
of the original Hamiltonian \eqref{Hamiltonian} or \eqref{Baxter} 
can be obtained by taking the homogeneous limit
$\xi_m\to\eta/2$.

To apply the algebraic Bethe ansatz to the computation of
\eqref{correlation}, we must express local 
fermionic operators in terms of  elements of the monodromy 
operators \eqref{monodromy}. 
In fact, such an expression has already been obtained by 
solving the quantum inverse scattering problem for local fermion 
operators \cite{GK00}.

\begin{theorem} \label{inverse} \cite{GK00}
For the inhomogeneous spinless fermion model \eqref{monodromy} with 
arbitrary inhomogeneous parameters $\xi_{m}$ ($m=1,\dots  M$), 
the local fermionic operators acting on the  $j$th space can 
be expressed in terms of the elements of the fermionic 
monodromy operator as
\begin{alignat}{2}
&c_{j}^{\dagger}=\prod_{k=1}^{j-1}\mathcal{T}(\xi_{k})
                         B(\xi_{j})\prod_{k=1}^{j}\mathcal{T}^{-1}(\xi_{k}),
&\qquad& c_{j}=\prod_{k=1}^{j-1}\mathcal{T}(\xi_{k})C(\xi_{j})
                    \prod_{k=1}^{j}\mathcal{T}^{-1}(\xi_{k}),  \nn \\
&n_{j}=-\prod_{k=1}^{j-1}\mathcal{T}(\xi_{k})D(\xi_{j})
                   \prod_{k=1}^{j}\mathcal{T}^{-1}(\xi_{k}),
&\qquad& \bm{1}=\prod_{k=1}^M \mathcal{T}(\xi_k).
\label{qip}
\end{alignat}
\end{theorem}

To proceed the evaluation by use of the theorem~\ref{inverse}, one
needs to calculate the action of the operators $A$, $B$,
$C$ and $D$ on an arbitrary state
$\bra \psi |=\bra \Omega|\prod_{j=1}^{N}C(\lambda_{j})$,
where $\{\lambda\}$ is a set of  arbitrary 
complex numbers (not necessary the Bethe  roots). 
After simple but tedious computation utilizing \eqref{commutation},
one obtains
\begin{align}
\bra \Omega|\prod_{j=1}^{N}C(\lambda_{j})A(\lambda_{N+1})=&
         \sum_{b=1}^{N+1}a(\lambda_{b})\frac{\prod_{j=1}^{N}
                 \sh(\lambda_{j}-\lambda_{b}+\eta)}
                   {\prod_{\substack{j=1 \\ j \neq b}}^{N+1}
                         \sh(\lambda_{j}-\lambda_{b})}
             \bra \Omega| \prod_{\substack{j=1 \\ j \neq b}}^{N+1}
                                                 C(\lambda_{j}), \nn \\
\bra \Omega|\prod_{j=1}^{N}C(\lambda_{j})D(\lambda_{N+1})=&
         (-1)^{N}\sum_{a=1}^{N+1}d(\lambda_{a})\frac{\prod_{j=1}^{N}
                 \sh(\lambda_{a}-\lambda_{j}+\eta)}
                   {\prod_{\substack{j=1 \\  j \neq a}}^{N+1}
                  \sh(\lambda_{a}-\lambda_{j})}
             \bra \Omega| \prod_{\substack{j=1 \\ j \neq a}}^{N+1}
                                                 C(\lambda_{j}), \nn \\
\bra \Omega|\prod_{j=1}^{N}C(\lambda_{j})B(\lambda_{N+1})=&
         (-1)^{N-1}\sum_{a=1}^{N+1}d(\lambda_{a})\frac{\prod_{k=1}^{N}
                 \sh(\lambda_{a}-\lambda_{k}+\eta)}
                {\prod_{\substack{k=1 \\  k \neq a}}^{N+1}
                 \sh(\lambda_{a}-\lambda_{k})} \nn \\
 \times \sum_{\substack{a'=1\\a'\neq a}}^{N+1}
     & \frac{a(\lambda_{a'})}{\sh(\lambda_{N+1}-\lambda_{a'}+\eta)}
     \frac{\prod_{\substack{j=1 \\ j\neq a}}^{N+1}
                      \sh(\lambda_{j}-\lambda_{a^{\prime}}+\eta)}
           {\prod_{\substack{j=1 \\  j \neq a,a^{\prime}}}^{N+1}
                    \sh(\lambda_{j}-\lambda_{a^{\prime}})} 
\bra \Omega| \prod_{\substack{j=1 \\ j \neq a,a^{\prime}}}^{N+1}
                                          C(\lambda_{j}).
\label{action}
\end{align}
Note that the extra sign factors $(-1)^N$ and $(-1)^{N+1}$
appearing in the last two relations reflect the fermionic
nature of the system (cf. (3.7) and (3.8) in \cite{KMST02}).
Finally the action of the operator $C(\lambda)$ is free.

In the above, one distinguishes   the terms containing $a(\lambda_{N+1})$
or $d(\lambda_{N+1})$ and the others, which are referred to
as the direct and indirect terms, respectively.
Due to the existence of indirect terms, the explicit
form of the scalar product 
$\bra \Omega| \prod_{j=1}^N C(\mu_j)\prod_{j=1}^N
     B(\lambda_j)|\Omega \ket$ is required to compute
the expectation value \eqref{correlation}, where
$\{\lambda_j\}_{j=1}^N$ are solutions of the Bethe
ansatz equation and $\{\mu_j\}_{j=1}^N$ are arbitrary
complex numbers.
This can be explicitly evaluated in the same way 
as in \cite{KBIbook,Slavnov}.
The difference between the present system and the XXZ chain
stems from the difference of the commutation relations 
\eqref{commutation}.
Here we  write down only the result. 
\
\begin{proposition} \label{scalar}
The scalar product between a Bethe state and an 
arbitrary state
\begin{equation}
\mathbb{S}_N(\{\mu\}|\{\lambda\})
=\bra \Omega| \prod_{j=1}^{N}C(\mu_{j})\prod_{j=1}^{N}
B(\lambda_{j})|\Omega \ket
\end{equation}
can be expressed  as follows:
\begin{equation}
\mathbb{S}_N(\{\mu\}|\{\lambda\})=\mathbb{S}_N(\{\lambda\}|\{\mu\})
 =(-1)^{\frac{N(N-1)}{2}}\frac{\prod_{a,b=1}^{N}\sh(\lambda_{a}-\mu_{b}+\eta)}
       {\prod_{a<b}^{N}\sh(\lambda_{a}-\lambda_{b})\sh(\mu_{b}-\mu_{a})}
  {\det}_{N}\Psi( \{ \mu \}| \{ \lambda \}),
\label{sp}
\end{equation}
where $\{ \lambda_{j} \}_{j=1}^N$  are Bethe roots, 
$\{ \mu_{j} \}_{j=1}^N $ are arbitrary complex parameters.
The $N \times N$ matrix $\Psi(\{ \mu \} | \{ \lambda \} )$ 
is defined by       
\begin{eqnarray}
\Psi_{jk}( \{ \mu \}| \{ \lambda \})=
  t(\lambda_{j},\mu_{k})-(-1)^{N+1}d(\mu_{k})t(\mu_{k},\lambda_{j})
  \prod_{a=1}^{N}   
      \frac{\sh(\mu_{k}-\lambda_{a}+\eta)}{\sh(\mu_{k}-\lambda_{a}-\eta)},
\end{eqnarray}
and ${\det}_N$ denotes the determinant of an $N\times N$ matrix.
\end{proposition}
Reflecting the fermion statistics, the extra factor $(-1)^{N+1}$
enters into the above expression.
This factor, however, is eliminated by that in the Bethe ansatz
equation \eqref{BAE}, when one takes the limit $\{\mu\}\to\{\lambda\}$,
which gives the square of the norm of the Bethe vector:
\begin{equation}
\mathbb{S}_N(\{\lambda\}|\{\lambda\})
  =(-1)^{\frac{N(N-1)}{2}}\sh^{N}(\eta) \prod_{\substack{a,b=1 \\ a \neq b}}^{N}
    \frac{\sh(\lambda_{a}-\lambda_{b}+\eta)}{\sh(\lambda_{a}-\lambda_{b})}
    {\det}_{N}\Phi( \{ \lambda \}),
\label{norm}
\end{equation}
where
\begin{equation}
\Phi_{jk}( \{ \lambda \})=\delta_{jk} 
          \left[ \frac{d^{\prime}(\lambda_{j})}
                 {d(\lambda_{j})}-\sum_{a=1}^{N}K(\lambda_{j}-\lambda_{a}) 
          \right]+K(\lambda_{j}-\lambda_{k}).
\end{equation}

Set $ \{ \mu \}=\{ \xi_{1},\dots,\xi_{n} \} 
    \cup \{ \lambda_{n+1},\dots, \lambda_{N} \} $ in \eqref{sp}. Then
the ratio of the determinants of $ \Psi $ and 
$ \Phi $ can be evaluated in the thermodynamic limit:
\begin{equation}
\frac{{\det}_{N}\Psi}{{\det}_{N}\Phi}=
   \prod_{a=1}^{n}(M\rho_{\rm tot}(\lambda_{a},\{\xi\}))^{-1}
    {\det}_{n}\rho(\lambda_{j},\xi_{k}),
\end{equation}
where $\rho(\lambda,\xi)$ and 
$\rho_{\rm tot}(\lambda,\{\xi\})$ are defined
as \eqref{inhom-density} and \eqref{total-density}, respectively.

%
\subsection{Form factors}
%
Utilizing the scalar product obtained in
the preceding subsection, we derive  determinant 
representations of the form factors for a finite 
system. 
These formulae might be useful in calculating
the spectral functions of the present model.

The form factors of the local fermion operators
are defined by
\begin{align}
&F_{N}^{-}(m| \{ \mu \}| \{ \lambda \})=
     \bra \Omega |\prod_{j=1}^{N+1}C(\mu_{j})c_{m}^{\dagger}
                  \prod_{k=1}^{N}B(\lambda_{k})
          |\Omega \ket, \nn \\
&F_{N}^{+}(m| \{ \lambda \}|\{ \mu \})=
     \bra \Omega |\prod_{k=1}^{N}C(\lambda_{k})c_{m}
                  \prod_{j=1}^{N+1}B(\mu_{j})|\Omega \ket, \nn \\
&
F_{N}^{z}(m| \{ \mu \}| \{ \lambda \})=
     \bra \Omega |\prod_{j=1}^{N}C(\mu_{j})(1-2n_{m})
                  \prod_{k=1}^{N}B(\lambda_{k})
          |\Omega \ket,
\end{align}
where $\{ \lambda_{k} \}_{k=1}^N$ and $\{ \mu_{j} \}_{j=1}^{N+1}$ (or 
$\{ \mu_{j} \}_{j=1}^{N}$) are solutions of the Bethe ansatz equation \eqref{BAE}.
To evaluate them, first we substitute  solutions of the
quantum inverse scattering problem for local fermion operators \eqref{qip}
into the above expressions. 
For instance, $F_N^z(m|\{\mu\}|\{\lambda\})$ can be written as
\begin{equation}
F_N^z(m|\{\mu\}|\{\lambda\})
=2\bra \Omega| \prod_{j=1}^N C(\mu_j) 
          \prod_{l=1}^{m-1} \mathcal{T}(\xi_l) A(\xi_m)
          \prod_{l=m+1}^{M}\mathcal{T}(\xi_l) 
          \prod_{k=1}^NB(\lambda_k)|\Omega \ket-
       \mathbb{S}_N(\{\mu\}|\{\lambda\}).
\end{equation}
Then, noting that the states $\prod B(\lambda_k)|\Omega\ket$ 
and $\bra \Omega|\prod C(\mu_j)$ are the Bethe states, we remove 
the products of the transfer matrices by \eqref{eigen}.
Finally applying the determinant representation 
of the scalar product \eqref{sp}, and using  a 
simple identity  $\prod_{j=1}^N \prod_{k=1}^{M}
f(\lambda_{j},\xi_{k})=1$ derived by the Bethe ansatz 
equation \eqref{BAE},
we obtain the determinant representations of the form factors.
They are summarized into the following proposition.
\begin{proposition}
The form factors $F_N^-(m|\{\mu\}|\{\lambda\})$ 
and $F_N^+(m|\{\lambda \}|\{\mu\})$ for local fermion
operators $c_m^{\dagger}$ and $c_m$ are respectively
given by the following determinant representations:
\begin{align}
F_{N}^{-}(m| \{ \mu \}| \{ \lambda \})=&
   (-1)^{\frac{N(N+1)}{2}} 
      \frac{\phi^{N+1}_{m-1}(\{ \mu \})}{\phi^N_{m-1}(\{ \lambda \})}
     \frac{\prod_{j=1}^{N+1}\sh(\mu_{j}-\xi_{m}+\eta)}
          {\prod_{k=1}^{N}\sh(\lambda_{k}-\xi_{m}+\eta)} \nn \\
&\quad
\times 
\frac{\det_{N+1}H^{-}(m| \{ \mu \}| \{ \lambda \})}
         {\prod_{j < k}^{N+1}\sh (\mu_{k}-\mu_{j})
         \prod_{\alpha < \beta }^{N}
              \sh (\lambda_{\alpha}-\lambda_{\beta})}, \nn \\
F_{N}^{+}(m| \{ \lambda \}|\{ \mu \})=&
      \frac{\phi_{m}^N(\{ \lambda \})\phi_{m-1}^N(\{ \lambda \})}
           {\phi_{m-1}^{N+1}(\{ \mu \})\phi_{m}^{N+1}(\{ \mu \})}
                   F_{N}^{-}(m|  \{ \mu \}|\{ \lambda \}),
\label{fp}
\end{align}
where the coefficient $\phi_{m}^N(\{ \lambda \})$ is
\begin{equation}
\phi_{m}^N(\{ \lambda \})=\prod_{j=1}^{N}\prod_{k=1}^{m}f(\lambda_{j}, \xi_{k}),
\end{equation}
and the $(N+1)\times (N+1)$ matrix $H^{-}(m|\{ \mu \}| \{ \lambda \})$ 
is defined by
\begin{align}
&H_{jk}^{-}=\frac{\sh\eta}{\sh (\mu_{j}-\lambda_{k})}
     \left[ \prod_{\substack{a=1 \\a \neq j}}^{N+1}\sh (\mu_{a}-\lambda_{k}+\eta)
        -(-1)^N d(\lambda_{k}) 
            \prod_{\substack{a=1 \\a \neq j}}^{N+1} 
                        \sh (\mu_{a}-\lambda_{k}-\eta)  \right] 
\text{ for $\ k \le N$}, \nn \\
&H_{jN+1}^{-}=t(\mu_j,\xi_m).
\label{Hp}
\end{align}
On the other hand, the form factor $F_N^z(m|\{\mu\}|\{\lambda\})$ for the
operator $1-2n_m$ is given by
\begin{align}
F_{N}^{z}(m| \{ \mu \}| \{ \lambda \})=& 
   (-1)^{\frac{N(N-1)}{2}} \frac{\phi_{m-1}^N(\{ \mu \})}{\phi_{m-1}^N(\{ \lambda \})}
      \prod_{j=1}^{N}\frac{\sh (\mu_{j}-\xi_{m}+\eta)}
                          {\sh (\lambda_{j}-\xi_{m}+\eta)} \nn \\
& \quad \times
\frac{\det_{N}(H(\{ \mu\}|\{ \lambda\} \})-2P(m| \{ \mu \}| \{ \lambda \}))}
             {\prod_{j<k}^N[\sh(\lambda_j-\lambda_k)\sh (\mu_{k}-\mu_{j})]},
\label{fz}
\end{align}
where $H(\{ \mu \}| \{ \lambda \})$ is an $N \times N$ matrix defined as
\begin{align}
H_{jk}=\frac{\sh \eta}{\sh (\mu_{j}-\lambda_{k})} 
          \left[\prod_{a \neq j}\sh (\mu_{a}-\lambda_{k}+\eta)
                    -(-1)^{N+1}d(\lambda_{k})\prod_{a \neq j}
                           \sh (\mu_{a}-\lambda_{k}-\eta)\right],
\label{Hz}
\end{align}
and $P(m| \{ \mu \}| \{ \lambda \})$ is an $N\times N$ matrix 
of rank one, 
\begin{align}
P_{jk}(m)=t(\mu_j,\xi_m)\prod_{a=1}^{N}\sh (\lambda_{a}-\lambda_{k}+\eta).
\end{align}

\end{proposition}
Here we have set $a(\lambda)=1$.
Note that, in the derivation of $F_N^z(m|\{\mu\}|\{\lambda\})$,
we have used the orthogonality of the Bethe vectors, and the
formula that the determinant of the sum of an arbitrary 
$N\times N$ matrix $\mathcal{A}$ and an $N\times N$ matrix of
rank one $\mathcal{B}$ can be given by
\begin{equation}
\det(\mathcal{A}+\mathcal{B})=\det\mathcal{A}+\sum_{l=1}^N
                              \det\mathcal{A}^{(l)},
\quad
\mathcal{A}_{jk}^{(l)}=\begin{cases}
                        \mathcal{A}_{jk} &\text{ for $k\neq l$} \\
                        \mathcal{B}_{jk} &\text{ for $k=l$}
                     \end{cases}.
\end{equation}

Comparing  the form factor for the local spin operator
of the third component $\sigma_m^z$ in the XXZ chain \cite{KMT99}, 
one finds that the extra factor $(-1)^{N+1}$ appears in 
\eqref{Hz}. 
This factor, however, is canceled by that in the Bethe 
ansatz equation \eqref{BAE}.
Hence, in the thermodynamic limit where the distribution
of the Bethe roots \eqref{distribution} characterizing the 
ground state is the same as that of the XXZ chain, 
$F_N^z(m|\{\mu\}|\{\lambda\})$ \eqref{fz}  coincides with the 
form factor of the $\sigma_m^z$ for the XXZ chain\footnote{
Note that the overall factor $(-1)^{N(N-1)/2}$ in \eqref{fz} and
\eqref{fp} can
be eliminated by the normalization of the Bethe vector 
(see \eqref{norm}).}.
In contrast to this, for the form factor $F_N^{\pm}(m|\{\mu\}|\{\lambda\})$
\eqref{fp}, the factor $(-1)^N$ in \eqref{Hp} can not be eliminated by the
Bethe ansatz equation \eqref{BAE}. Namely, compared with the form factors of 
the $\sigma_m^{\pm}$ for the XXZ chain, the extra minus sign remains in \eqref{Hp}.
This essential difference caused by the statistics between the fermion 
and the spin induces the different behavior of the correlation 
functions $\bra c_1c_{m+1}^{\dagger} \ket$ and 
$\bra \sigma_1^+\sigma_{m+1}^-\ket$.

%
\section{Multiple integral representations for correlation functions}
%
In this section, we derive  multiple integrals representing 
the one-particle Green's function and the density-density 
correlation function.
Inserting the expression of the local fermion operators
$c_j^{\dagger}$ and $c_j$ \eqref{qip} into \eqref{correlation}, 
one obtains 
\begin{equation}
\bra c_1 c_{m+1}^{\dagger} \ket=
   \frac{\bra \psi_{\rm g}|C(\xi_1)\prod_{a=2}^m (A-D)(\xi_a) 
                           B(\xi_{m+1})\prod_{b=1}^{m+1} 
                           (A-D)^{-1}(\xi_b) 
        |\psi_{\rm g}\ket}{\bra \psi_{\rm g}| \psi_{\rm g}\ket}.
\label{green}
\end{equation}
Note that $\bra c_1^{\dagger} c_{m+1}\ket=-\bra c_1 c_{m+1}^{\dagger}\ket$.

Following \cite{KMST02}, for the density-density correlation function, 
we conveniently introduce an operator $Q_{1,m}$ as $Q_{1,m}=\sum_{k=1}^m n_k$
and consider the expectation value of the 
generating function $\exp(\beta Q_{1,m})$ 
($\beta\in\mathbb{C}$):
\begin{equation}
\bra \exp(\beta Q_{1,m})\ket=
\frac{\bra \psi_{\rm g}|\prod_{a=1}^m (A-\e^{\beta}D)(\xi_a) 
                           \prod_{b=1}^{m} 
                           (A-D)^{-1}(\xi_b) 
        |\psi_{\rm g}\ket}{\bra \psi_{\rm g}| \psi_{\rm g}\ket}.
\label{density1}
\end{equation}
Then the density-density correlation  can be 
expressed as
\begin{equation}
\bra n_1 n_{m+1}\ket=\frac{1}{2}\mathcal{D}_m^2
       \frac{\partial^2}{\partial \beta^2}
              \bra \exp(\beta Q_{1,m})\ket
              \Bigr|_{\beta=0},
\label{density-density}
\end{equation}
where $\mathcal{D}_m$ denotes the lattice derivative defined
by
$\mathcal{D}_m f(m)=f(m+1)-f(m)$ and $\mathcal{D}_m^2 f(m)=
f(m+1)-2f(m)+f(m-1)$. 
In fact, there exist several manners to express the density-density correlation
by the generating function \cite{KMST02}.
For instance, $\bra n_1 n_{m+1} \ket$ is also written by using 
$\bra \exp(\beta Q_{1,m}) n_{m+1} \ket$ as
\begin{align}
&\bra n_1 n_{m+1} \ket=\mathcal{D}_{m-1}
       \frac{\partial}{\partial \beta}
              \bra \exp(\beta Q_{1,m})n_{m+1}\ket
              \Bigr|_{\beta=0}, \nn \\
& 
\bra \exp(\beta Q_{1,m})n_{m+1}\ket
  =-\frac{\bra \psi_{\rm g}|\prod_{a=1}^m (A-\e^{\beta}D)(\xi_a) 
                           D(\xi_{m+1})
                           \prod_{b=1}^{m+1} 
                           (A-D)^{-1}(\xi_b) 
        |\psi_{\rm g}\ket}{\bra \psi_{\rm g}| \psi_{\rm g}\ket}.
\label{density2}
\end{align}
%

%
%

To evaluate the correlation functions \eqref{green}--\eqref{density2},
we must calculate the multiple action of $(A-\e^{\beta}D)(x)$ on an
arbitrary state $\bra \psi|=\bra \Omega| \prod_{j=1}^N C(\lambda_j)$,
where $\{\lambda\}$ are arbitrary complex numbers.
Utilizing the commutation relations among $A$, $D$ and $C$ \eqref{commutation},
one finds that the action on a general state can be written as
\begin{align}
\bra \psi|\prod_{a=1}^m &(A-\e^{\beta}D)(x_a) \nn \\ 
  &
=
\sum_{n=0}^p \sum_{\substack{\{\lambda\}=\{\lambda^+\}\cup\{\lambda^-\}\\
                             \{x\}=\{x^+\}\cup\{x^-\}\\
                             |\lambda^+|=|x^+|=n}}
          R_n(\{x^+\}|\{x^-\}|\{\lambda^+\}|\{\lambda^-\})
         \bra \Omega|\prod_{a=1}^{n} C(x^+_a)
                     \prod_{b=1}^{N-n}C(\lambda_b^-),
\label{mult}
\end{align}
where $|\lambda^+ |$, $|x^+|$, etc denote the number of elements of
$\{\lambda^+\}$, $\{x^+\}$, etc; $p=\min(m,N)$.

\begin{proposition}\label{prop-mult}
The coefficient  $R_n(\{x^+\}|\{x^-\}|\{\lambda^+\}|\{\lambda^-\})$
in \eqref{mult} is given by
\begin{align}
R_n(\{x^+\}|\{x^-\}|\{\lambda^+\}|\{\lambda^-\})=&
     S_{n}(\{x^+\}|\{\lambda^+\}|\{\lambda^-\}) 
       \prod_{a=1}^{m-n}\biggl[a(x^-_{a})\prod_{b=1}^{n}f(x^+_{b},x^-_{a})
       \prod_{b=1}^{N-n}f(\lambda^-_{b},x^-_{a}) \nn \\
&
 -\e^{\beta}d(x^-_{a})\prod_{b=1}^{n}\{-f(x^-_{a},x^+_{b}) \} 
                    \prod_{b=1}^{N-n} \{-f(x^-_{a},\lambda^-_{b}) \} \biggr],
\end{align}
where the highest coefficient is expressed as
\begin{equation}
S_{n}( \{ x^+ \}| \{ \lambda^+ \}| \{ \lambda^- \})=
                  \frac{\prod_{a,b=1}^{n} 
                        \sh  (x^+_a-\lambda^+_b+\eta)}
                   {\prod_{a<b}^{n}\left[\sh(\lambda^+_b-\lambda^+_a)
                    \sh(x^+_a-x^+_b)\right]}{\det}_n M_{jk}
\end{equation}
with
\begin{align}
M_{jk}=&a(\lambda^+_{j})t(x^+_{k},\lambda^+_{j})\prod_{a=1}^{N-n}
                    f(\lambda^-_{a},\lambda^+_{j})              \nn \\
&+\e^{\beta}d(\lambda^+_{j})t(\lambda^+_{j},x^+_{k})
       \prod_{a=1}^{N-n}\{-f(\lambda^+_{j},\lambda^-_{a}) \}
       \prod_{b=1}^{n} \left\{-\frac{\sh(\lambda^+_{j}-x^+_{b}+\eta)}
                           {\sh  (\lambda^+_{j}-x^+_{b}-\eta)} \right\}.
\end{align}
\end{proposition}
The proof is completely parallel to that in the XXZ chain \cite{KMST02}.
The difference of the expression stems from the difference of the
commutation relation of $C$ and $D$, which reflects the fermionic
nature of the present system.

If $\bra \psi |$ is a Bethe vector (or equivalently the eigenstate of the
transfer operator \eqref{TM}), one further has the following expression
which is directly applicable to the computation of the generating function
\eqref{density1}.
\begin{corollary}\label{cor-mult}
If $\bra \psi|=\bra \Omega|\prod_{j=1}^NC(\lambda_j)$ is a
Bethe vector, namely if $\{\lambda\}$ are the solutions to
the Bethe ansatz equation \eqref{BAE}, the coefficient 
$R_n(\{\xi^+\}|\{\xi^-\}|\{\lambda^+\}|\{\lambda^-\})$ of the  action 
$\prod_{a=1}^m(A-\e^{\beta}D)(\xi_a)$ is given by
\begin{align}
&R_n(\{\xi^+\}|\{\xi^-\}|\{\lambda^+\}|\{\lambda^-\})=
                 \widetilde{S}_{n}(\{ \xi^+ \}|\{ \lambda^+ \}) \nn \\
&\qquad \qquad \qquad \times   \prod_{a=1}^{n}\prod_{b=1}^{m-n}f(\xi^+_a,\xi^-_b)
   \prod_{a=1}^{N-n}\prod_{b=1}^{m-n}f(\lambda^-_a,\xi^-_b)
   \prod_{a=1}^{N-n}\prod_{b=1}^{n}f(\lambda^-_a,\lambda^+_b),
\end{align}
where
\begin{equation}
\widetilde{S}_{n}(\{\xi^+\}|\{\lambda^+\})=
           \frac{\prod_{a,b=1}^{n}
                           \sh(\xi^+_a-\lambda^+_b+\eta)}
                {\prod_{a<b}^{n}
                      \left[
                      \sh(\lambda^+_b-\lambda^+_a)
                      \sh(\xi^+_{a}-\xi^+_{b})   
                       \right]}
                 {\det}_n\widetilde{M}_{jk}.
\end{equation}
Here the $n\times n$ matrix $\widetilde{M}$ is defined as
\begin{align}
\widetilde{M}_{jk}=t(\xi^+_{k},\lambda^+_{j})+
            \e^{\beta}t(\lambda^+_{j},\xi^+_{k})
          \prod_{a=1}^{n}\left[\frac{\sh(\lambda^+_{a}-\lambda^+_{j}+\eta)}
                               {\sh(\lambda^+_{j}-\lambda^+_{a}+\eta)}
                         \frac{\sh(\lambda^+_{j}-\xi^+_{a}+\eta)}
                                {\sh(\xi^+_{a}-\lambda^+_{j}+\eta)}\right].
\end{align}
\end{corollary}
Note that here we have set $a(\lambda)=1$ and $d(\xi_a)=0$, and used the fact that
the parameters $\{\lambda\}$ satisfy the Bethe ansatz equation \eqref{BAE}.
%
%

Now we would like to compute the correlation functions.
First let us consider the generating function \eqref{density1} of the
density-density correlation $\bra n_1n_{m+1} \ket$.
Noting that $|\psi_{\rm g} \ket$ is the Bethe vector
characterizing the ground state, one finds that the numerator 
in \eqref{density1} can be expressed by 
$\bra \psi_{\rm g}|
    \prod_{a=1}^m(A-\e^{\beta}D)(\xi_a) 
        |\psi_{\rm g}\ket/\prod_{a=1}^m\tau(\xi_a)$, where $\tau(\lambda)$ is the corresponding
eigenvalue of the transfer matrix (see \eqref{eigen}). 
Then from Corollary~\ref{cor-mult}, Proposition~\ref{scalar} and the
norm of the Bethe vector \eqref{norm}, the generating function
\eqref{density1} is given by a determinant form. 
By $d(\xi_a)=0$, it immediately follows
that the determinant form is the same  as that
for the generating function of the longitudinal spin-spin
correlation function $\bra \sigma_1^z \sigma_{m+1}^z \ket$ 
\cite{KMST02}. 
Thus utilizing the method proposed in \cite{KMST02}, 
we arrive at the following proposition.
\begin{proposition} \label{gen-density1}
In the thermodynamic limit $M\to\infty$,
the ground state expectation value of the
generating function $\exp(\beta Q_{1,m})$ is expressed as  
\begin{align}
\bra \exp(\beta Q_{1,m}) \ket =&\sum_{n=0}^{m}\frac{1}{(n!)^{2}}
      \oint_{\Gamma}\prod_{j=1}^{n}\frac{\d z_{j}}{2\pi i}
       \int_{\mathcal{C}}\d^{n}\lambda 
        \prod_{a=1}^{n}\prod_{b=1}^{m}\frac{\sh(z_a-\xi_b+\eta)\sh(\lambda_{a}-\xi_{b})}
                                           {\sh(z_a-\xi_b)\sh(\lambda_a-\xi_b+\eta)} 
                                                                       \nn \\
&\times
       W_{n}( \{ \lambda \}| \{ z\}) 
        {\det}_{n}[\widetilde{M}_{jk}(\{ \lambda \}|\{ z \})] 
          {\det}_{n} [ \rho(\lambda_{j},z_{k}) ],
\label{mult-density1}
\end{align}
where
\begin{align}
W_{n}( \{ \lambda \}| \{ z \})=\prod_{a=1}^{n}\prod_{b=1}^{n}
        \frac{\sh(\lambda_{a}-z_{b}+\eta)\sh(z_{b}-\lambda_{a}+\eta)}
             {\sh(\lambda_{a}-\lambda_{b}+\eta)\sh(z_{a}-z_{b}+\eta)},
\end{align}
and
\begin{eqnarray}
\widetilde{M}_{jk}( \{ \lambda \}|\{ z \})=
         t(z_{k}, \lambda_{j})+\e^{\beta}t(\lambda_{j},z_{k})
     \prod_{a=1}^{n}\frac{\sh(\lambda_{a}-\lambda_{j}+\eta)
                           \sh(\lambda_{j}-z_{a}+\eta)}
                    {\sh(\lambda_{j}-\lambda_{a}+\eta)
                     \sh(z_{a}-\lambda_{j}+\eta)}.
\label{mat-M}
\end{eqnarray}
The contour $\mathcal{C}$ depends on the regime and
the magnitude of the chemical potential, while $\Gamma$ 
encircles $\{\xi_j\}_{j=1}^m$ and does not
contain any other singularities. 
In the above expression, the homogeneous limit $\xi_{j}=\eta/2$
can be taken trivially. 
In this case,  $\Gamma$ surrounds the point $\eta/2$. 
\end{proposition}
Note that here to replace the sums over the partition of 
the set $\{\xi\}$ with the set of contour integrals
around $\{\xi\}$, we have used the following relation:
\begin{align}
\sum_{\substack{\{\xi\}=\{\xi^+ \}\cup\{\xi^-\}\\
                |\xi^+|=n}}&
  \prod_{a=1}^n \prod_{b=1}^{m-n}f(\xi_a^+,\xi_b^-)\mathcal{F}(\{\xi^+\}) \nn \\
&=
\frac{1}{n!}\oint_{\Gamma} \prod_{j=1}^n\frac{\d z_j}{2\pi\i}
             \prod_{a=1}^n\prod_{b=1}^mf(z_a,\xi_b)
     \frac{\prod_{a=1}^n\prod_{\substack{b=1\\ b\neq a}}^n
           \sh(z_a-z_b)}
           {\prod_{a=1}^n\prod_{b=1}^n\sh(z_a-z_b+\eta)}\mathcal{F}(\{z\}),
\end{align}
where $\mathcal{F}(\{z\})$ is assumed to be a symmetric function of 
$n$ variables $\{z\}$, and to be analytic in the 
vicinities of $z_j=\xi_k$. 
As for the sum over the partition of the Bethe roots 
$\{\lambda\}$ \eqref{BAE} characterizing  the ground state, 
we have used the following 
relation, which is valid in the thermodynamic limit:
\begin{align}
\lim_{M\to\infty}\frac{1}{M^n}
   \sum_{\substack{\{\lambda\}=\{\lambda^+\}\cup\{\lambda^-\}\\
                   |\lambda^+|=n}}
            \mathcal{F}(\{\lambda^+\})
   =\frac{1}{n!}\int_{\mathcal{C}}\d^n \lambda
      \prod_{j=1}^n\[  \rho_{\rm tot}(\lambda_j,\{\xi\})\]
         \mathcal{F}(\{\lambda \}),
\end{align}
where $\mathcal{F}(\{\lambda\})$ is assumed to be a symmetric
function of $n$ variables $\{\lambda \}$, and to be zero when
any two of its variables are the same.

The multiple integral \eqref{mult-density1}
 representing the generating function
of the density-density correlation function is exactly the
same as that for the longitudinal spin-spin correlation function
\cite{KMST02}.
The density-density correlation function
$\bra n_1 n_{m+1} \ket$
can be obtained by taking the derivative with respect to $\beta$ 
and $m$ as in \eqref{density-density}.

As for the generating function 
$\bra \exp(\beta Q_{1,m})n_{m+1} \ket$ in \eqref{density2},
the numerator can be written as $\bra \psi_{\rm g}|
    \prod_{a=1}^m(A-\e^{\beta}D)(\xi_a) D(\xi_{m+1})
        |\psi_{\rm g}\ket/\prod_{a=1}^m\tau(\xi_a)$.
Applying Corollary~\ref{cor-mult}, and then 
using the second relation in \eqref{action},
one can obtain a determinant form of the
generating function. 
By the same method as in the derivation of 
\eqref{mult-density1}, the following multiple integral
representing $\bra \exp(\beta Q_{1,m})n_{m+1} \ket$ 
can be derived in the thermodynamic limit.
\begin{proposition}
The ground state expectation value of the operator 
$\bra \exp(\beta Q_{1,m})n_{m+1} \ket$ 
in the thermodynamic limit is written as
\begin{align}
\bra \exp(\beta Q_{1,m})n_{m+1} &\ket=\sum_{n=0}^{m}
   \frac{1}{(n!)^{2}}\oint_{\Gamma}\prod_{j=1}^{n}\frac{\d z_{j}}{2\pi i}
   \int_{\mathcal{C}}\d^{n+1}\lambda
     \prod_{a=1}^{n}\prod_{b=1}^{m}\frac{\sh(z_a-\xi_b+\eta)\sh(\lambda_{a}-\xi_{b})}
                                           {\sh(z_a-\xi_b)\sh(\lambda_a-\xi_b+\eta)} 
                                                    \nn \\
&
\times  \prod_{a=1}^{n}\frac{\sh(\lambda_{a}-\xi_{m+1})}{\sh(z_{a}-\xi_{m+1})}\prod_{a=1}^{n}\frac{\sh(\lambda_{n+1}-z_{a}+\eta)}
                           {\sh(\lambda_{n+1}-\lambda_{a}+\eta)}
  W_{n}( \{ \lambda \}| \{ z \})      \nn \\
&\times
{\det}_{n}[ \widetilde{M}_{jk}( \{ \lambda \}| \{ z \}) ]
{\det}_{n+1} [ \rho(\lambda_{j},z_{k}),\dots, \rho(\lambda_{j},z_{n}), 
                \rho(\lambda_{j},\xi_{m+1})].
\label{mult-density2}
\end{align}
Here the integration contours $\mathcal{C}$ and $\Gamma $ are 
taken as in Proposition~\ref{gen-density1}.
In this representation, the homogeneous limit $\xi_j=\eta/2$ can be 
trivially taken.
\end{proposition}
Note that the above expression agrees with that for 
$\bra \exp(\beta Q_{1,m}(1-\sigma_{m+1}^z)/2) \ket$ \cite{KMST02}, 
as expected.

Finally we would like to consider the equal-time one particle
Green's function $\bra c_1 c_{m+1}^{\dagger} \ket$ \eqref{green}.
Because the operators $B$ and $C$ enter into the numerator
in \eqref{green}, the computation is
more complicated than that for the generating functions
of the density-density correlation function. 
Nevertheless,   applying Proposition~\ref{prop-mult}
and then utilizing the third relation in \eqref{action},
one finds the method used above is still applicable 
for the evaluation of $\bra c_1c_{m+1}^{\dagger} \ket$.

\begin{proposition}\label{prop-mult-green}
The equal-time one-particle Green's function 
$\bra c_{1}c_{m+1}^{\dagger} \ket$ in the thermodynamic limit
can be represented as a multiple integral as
\begin{align}
\bra &c_{1}c_{m+1}^{\dagger} \ket = 
    -\sum_{n=0}^{m-1}\frac{1}{n!(n+1)!}
        \oint_{\Gamma}\prod_{j=1}^{n+1}\frac{\d z_{j}}{2\pi i}
        \int_{\mathcal{C}}\d^{n+1}\lambda  
        \int_{\widetilde{\mathcal{C}}}\d \lambda_{n+2}  
    \frac{ \prod_{a=1}^{n+1}\prod_{b=1}^m\sh(z_a-\xi_b+\eta)}
         { \prod_{a=1}^{n+1}\prod_{b=2}^{m+1}\sh(z_a-\xi_b)} \nn \\
&
\times \frac{\prod_{a=1}^n\prod_{b=2}^{m+1}\sh(\lambda_a-\xi_b)}
                         {\prod_{a=1}^n\prod_{b=1}^m\sh(\lambda_a-\xi_b+\eta)}
\frac{\prod_{a=1}^{n+1}[\sh(\lambda_{n+1}-z_a+\eta)\sh(\lambda_{n+2}-z_a-\eta)]}
     {\prod_{a=1}^n[\sh(\lambda_{n+1}-\lambda_a+\eta)
                                            \sh(\lambda_{n+2}-\lambda_a-\eta)]} 
                                                                  \nn \\
&\times \frac{ \widehat{W}_n(\{\lambda\}|\{z\})}
            {\sh(\lambda_{n+1}-\lambda_{n+2}+\eta)}
{\det}_{n+1}\widehat{M}_{jk}
{\det}_{n+2}[\rho(\lambda_j,z_1),\dots,\rho(\lambda_j,z_{n+1}),
                                            \rho(\lambda_j,\xi_{m+1})],
\label{mult-green}
\end{align}
where the function $\widehat{W}_n(\{\lambda\}|\{z\})$ and the
$(n+1)\times(n+1)$ matrix are respectively defined as
\begin{equation}
\widehat{W}_{n}(\{ \lambda \}| \{ z \})=
      \frac{\prod_{a=1}^{n}\prod_{b=1}^{n+1}
                      \sh(\lambda_{a}-z_{b}+\eta)\sh(z_{b}-\lambda_{a}+\eta)}
            {\prod_{a=1}^{n}\prod_{b=1}^{n}\sh(\lambda_{a}-\lambda_{b}+\eta)
             \prod_{a=1}^{n+1}\prod_{b=1}^{n+1}\sh(z_{a}-z_{b}+\eta)},
\end{equation}
and
\begin{equation}
\widehat{M}_{jk}= 
 \begin{cases}
     t(z_{k},\lambda_{j})+t(\lambda_j,z_{k})
        \prod_{a=1}^{n}\frac{\sh(\lambda_{a}-\lambda_{j}+\eta)}
                            {\sh(\lambda_{j}-\lambda_{a}+\eta)}
        \prod_{b=1}^{n+1}\frac{\sh(\lambda_{j}-z_{b}+\eta)}
                        {\sh(z_{b}-\lambda_{j}+\eta)} &\text{\quad for $j\le n$} \\
        t(z_k,\xi_1) &\text{\quad for $j=n+1$}
  \end{cases}.
\label{M}
\end{equation}
The integration contour $\mathcal{C}$ for the
variables $\{\lambda_j\}_{j=1}^{n+1}$ is taken as
in Proposition~\ref{gen-density1}, while the contour
$\widetilde{\mathcal{C}}$ for $\lambda_{n+2}$ should
be taken such that the contour $-\mathcal{C}\cup\widetilde{\mathcal{C}}$
surrounds the points $\lambda_{n+2}=z_1,\dots,z_{n+1}, \xi_{m+1}$
which are  simple poles of $\rho(\lambda_{n+2},z_k)$ and 
$\rho(\lambda_{n+2},\xi_{m+1})$, and does not
contain any other singularities. 
The homogeneous limit $\xi_j\to\eta/2$ can be taken trivially in the
above expression. Correspondingly the contours $\Gamma$
and $-\mathcal{C}\cup\widetilde{\mathcal{C}}$ encircle the point $\eta/2$.
\end{proposition}
%

%
\begin{figure}[ttt]
\begin{center}
\includegraphics[width=0.45\textwidth]{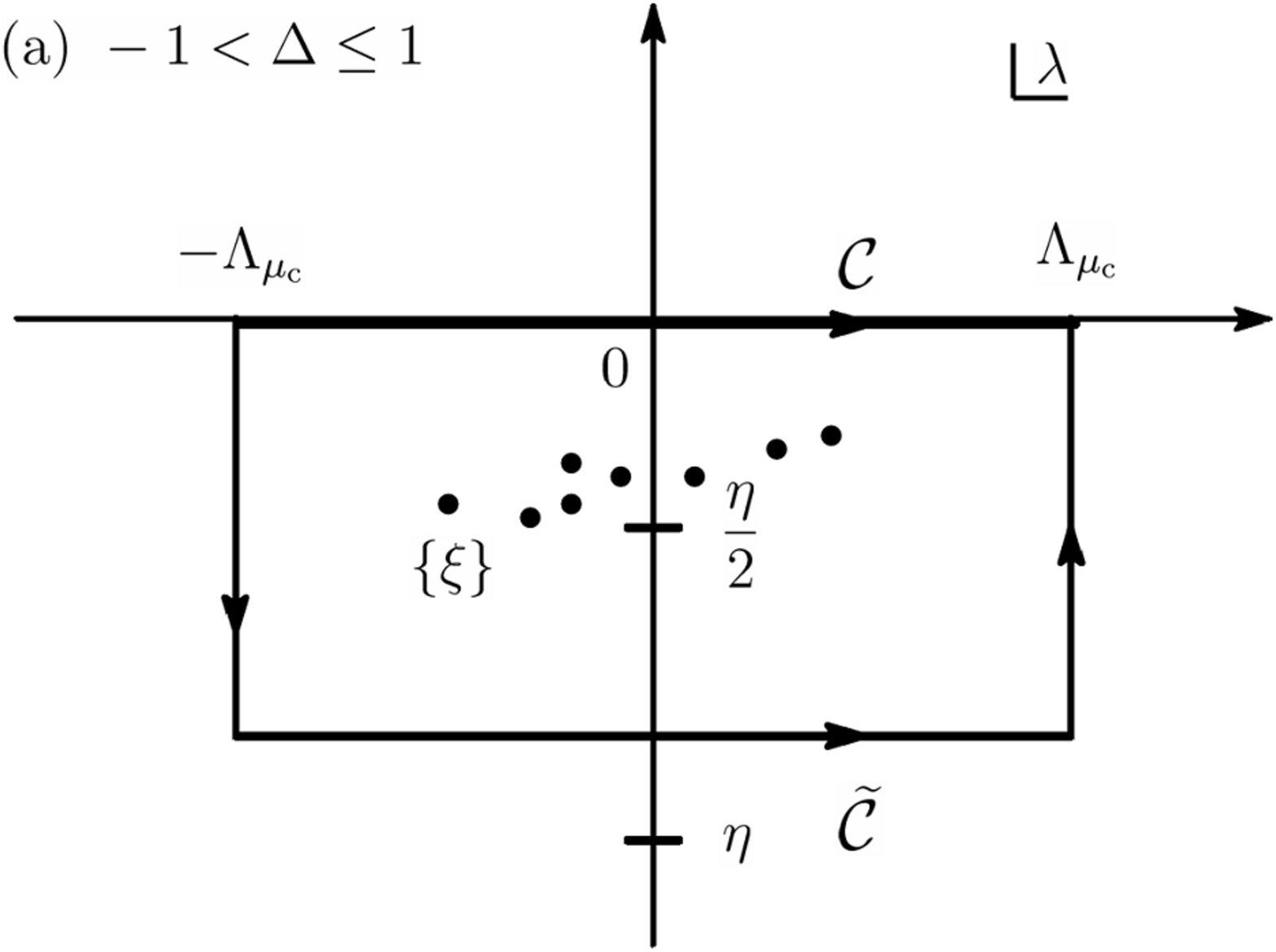}
\includegraphics[width=0.45\textwidth]{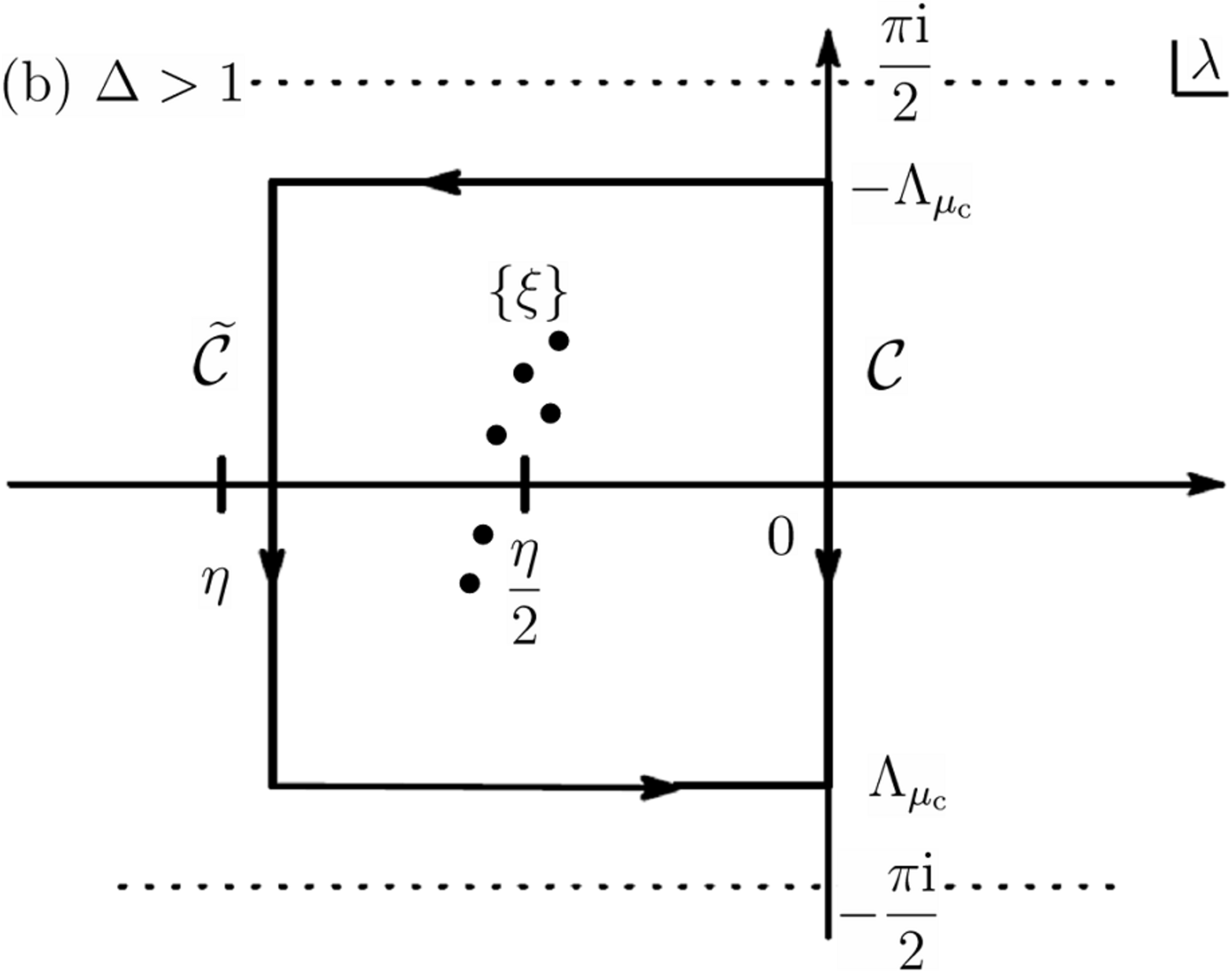}
\end{center}
\caption{The integration contours $\mathcal{C}$ and
$\widetilde{\mathcal{C}}$ for $-1<\Delta\le 1$ (a)
and for $\Delta>1$ (b).  The contour $\mathcal{-C}\cup
\widetilde{\mathcal{C}}$ encircles the singularities 
$\{\xi_j\}_{j=1}^m$ and does not
contain any other singularities.
}
\label{contour}
\end{figure}
%
%
In figure~\ref{contour}, typical examples of the integration contour 
$\widetilde{\mathcal{C}}$ are depicted both for the regimes
$-1<\Delta\le 1$ and $\Delta>1$.
Recalling that the Fermi point $\Lambda_{\mu_{\rm c}}$ at
zero chemical potential ($\mu_{\rm c}=0$) is 
$\Lambda_{\mu_{\rm c}}=\infty$ for $-1<\Delta\le 1$ 
and $\Lambda_{\mu_{\rm c}}=-\pi \i/2$ for $\Delta>1$ 
(see section~2), one finds that
the contributions from the right and left edges of 
$\widetilde{\mathcal{C}}$ disappear for $-1<\Delta\le 1$, 
and those from the  upper and the lower edges cancel each other 
for $\Delta>1$.
Using this together with the relation 
$\rho(\lambda+\eta,z)=-\rho(\lambda,z)$ which is valid 
only for $\mu_{\rm c}=0$, one arrives at the following
representation.
\begin{corollary}\label{prop-mult-green-hom}
The equal-time one-particle Green's function for the homogeneous
case at zero chemical potential can be expressed as
\begin{align}
\bra c_{1}  c_{m+1}^{\dagger}  \ket = &
    \sum_{n=0}^{m-1}\frac{1}{n!(n+1)!}
        \oint_{\Gamma}\prod_{j=1}^{n+1}\frac{\d z_{j}}{2\pi i}
        \int_{\mathcal{C}}\d^{n+2}\lambda  
   \prod_{a=1}^{n+1} \(\frac{ \sh(z_a+\frac{\eta}{2})}
                          { \sh(z_a-\frac{\eta}{2})}\)^m \nn \\
&
\times \prod_{a=1}^{n+1}\(\frac{\sh(\lambda_a-\frac{\eta}{2})}
                             {\sh(\lambda_a+\frac{\eta}{2})}\)^m
\frac{\prod_{a=1}^{n+1}[\sh(\lambda_{n+1}-z_a+\eta)\sh(\lambda_{n+2}-z_a)]}
     {\prod_{a=1}^n[\sh(\lambda_{n+1}-\lambda_a+\eta)
                                            \sh(\lambda_{n+2}-\lambda_a)]} 
                                                                  \nn \\
&\times \frac{ \widehat{W}_n(\{\lambda\}|\{z\})}
            {\sh(\lambda_{n+1}-\lambda_{n+2})}
{\det}_{n+1}\widehat{M}_{jk}
{\det}_{n+2}[\rho(\lambda_j,z_1),\dots,\rho(\lambda_j,z_{n+1}),
                                            \rho(\lambda_j,\frac{\eta}{2})],
\label{mult-green2}
\end{align}
where $\widehat{M}_{jk}$ is defined as \eqref{M} for $j\le n$
and $\widehat{M}_{jk}=t(z_k,\frac{\eta}{2})$ for $j=n+1$.
\end{corollary}
Compared with the transverse spin-spin correlation function
$\bra \sigma_1^+\sigma_{m+1}^-\ket$ in 
\cite{KMST02}\footnote{There exists a typo in 
equation (6.13) in \cite{KMST02} for the transverse spin-spin correlation
function $\bra \sigma_1^+ \sigma_{m+1}^-\ket$. It should
be corrected by multiplying by $-1$.
}, one finds the sign
in front of the second term of $\widehat{M}_{jk}$ for $j\le n$
changes.

%
\section{Free fermion model}
%
Using the multiple integral representations derived in 
the preceding section, we reproduce the exact expressions 
of correlation functions for the free fermion model.

Let us consider the density-density correlation
function $\bra n_1n_{m+1} \ket$ first.
As mentioned before, the generating functions
$\bra \exp(\beta Q_{1,m}) \ket$ \eqref{mult-density1}
for $\bra n_1n_{m+1} \ket$ is  the same as that for the
spin-spin correlation functions $\bra \sigma_1^z\sigma_{m+1}^z \ket$
for the XXZ chain. 
Hence the same method described in \cite{KMST-free} can be applied 
to derive the exact expression of $\bra n_1n_{m+1} \ket$.
Here we derive it by using \eqref{mult-density2} and \eqref{density2}.
Set $\eta=-\pi\i/2$.
Then $\widetilde{M}_{jk}$ \eqref{mat-M} in \eqref{mult-density2}
is factorized as
 $\widetilde{M}_{jk}=2(\e^{\beta}-1)/\sh(2(\lambda_j-z_k))$,
which significantly simplifies the integral representation.
After taking the derivative of \eqref{mult-density2} 
with respect to $\beta$ and setting $\beta=0$, one 
finds all the terms except for $n=1$ vanish.
Substituting $\rho(\lambda,z)=\i/(\pi\sh(2(\lambda-z))$,
and taking the lattice derivative, one obtains
\begin{equation}
\bra n_1n_{m+1}\ket=
\frac{\i}{2\pi^3}\oint_{\Gamma} \d z
\int_{\mathcal{C}}\d^2\lambda
\frac{\sh(\lambda_1-\lambda_2) \varphi^m(z) \varphi^{-m}(\lambda_1)}
     {\sh(\lambda_1-z)\sh(\lambda_2-z)\ch(2\lambda_1)\ch(2\lambda_2)},
\end{equation}
where 
\begin{equation}
\varphi(\lambda)=\frac{\sh(\lambda-\frac{\pi}{4}\i)}
                      {\sh(\lambda+\frac{\pi}{4}\i)}.
\end{equation}
The integral on $\Gamma$ can be evaluated by considering
the residues outside the contour $\Gamma$ i.e. at
the points $z=\lambda_1$ and $\lambda_2$.
It immediately follows that
\begin{equation}
\bra n_1n_{m+1}\ket=
\frac{1}{\pi^2}
\int_{\mathcal{C}}\d^2\lambda
\frac{1-\varphi^m(\lambda_2) \varphi^{-m}(\lambda_1)}
     {\ch(2\lambda_1)\ch(2\lambda_2)}.
\end{equation}
Changing the variables $\ch(2\lambda)=-1/\cos k$ 
($k\in [k_{\rm F},2\pi-k_{\rm F}]$), where
$k_{\rm F}\in[\pi/2,\pi]$ is the Fermi momentum
determined by $\cos k_{\rm F}=-1/\ch\Lambda_{\mu_{\rm c}}$
(see section 2), and identifying $\d \lambda=- \d k/(2\cos k)$, 
one arrives at
\begin{align}
\bra n_1 n_{m+1} \ket
&=\frac{1}{4\pi^2}\int_{k_{\rm F}}^{2\pi-k_{\rm F}} \d k_1
                 \int_{k_{\rm F}}^{2\pi-k_{\rm F}} \d k_2
      (1-\e^{-\i m (k_1-k_2)}) \nn \\
&=\bra n_j \ket^2-\frac{1-\cos (2 k_{\rm F} m)}{2 \pi^2 m^2}
\qquad k_{\rm F}\in [\frac{\pi}{2},\pi],
\end{align}
where $\bra n_j \ket=1-k_{\rm F}/\pi$ is the particle density
given by \eqref{particle-density}. This expression
agrees with the well-known result.
At zero chemical potential $\mu_{\rm c}=0$,
where the Fermi momentum corresponds to $k_{\rm F}=\pi/2$,
one finds
\begin{equation}
\bra n_1 n_{m+1} \ket=\frac{1}{4}-\frac{1-(-1)^m}{2\pi^2 m^2}.
\end{equation}

Next let us discuss the one-particle Green's function 
$\bra c_1 c_{m+1}^{\dagger}\ket$ \eqref{mult-green}.
Since $M_{jk}=0$ for $j\le n$,
only the term corresponding to $n=0$ survives 
in \eqref{mult-green}.
Namely
\begin{align}
\bra c_1 c_{m+1}^{\dagger}\ket
=&\frac{1}{2\pi^3}\oint_{\Gamma} \d z
  \int_{\mathcal{C}}\d \lambda_1
  \int_{\widetilde{\mathcal{C}}}\d \lambda_2
  \frac{\ch(\lambda_1-z)\ch(\lambda_2-z)}
       {\sh(z+\frac{\pi}{4}\i)\sh(z-\frac{\pi}{4}\i)}
  \frac{\varphi^m(z)}{\ch(\lambda_1-\lambda_2)} \nn \\
 & \times \[\frac{1}{\sh(2(\lambda_1-z))\sh(2(\lambda_2+\frac{\pi}{4}\i))}
           -\frac{1}{\sh(2(\lambda_2-z))\sh(2(\lambda_1+\frac{\pi}{4}\i))}\].
\end{align}
To evaluate the above integral explicitly, we shift the
$\lambda_2$-contour $\widetilde{\mathcal{C}}$ to $\mathcal{C}$. 
Since the integrand is antisymmetric with respect to 
the exchange of $\lambda_1$ and $\lambda_2$, the resulting
integral vanishes: we only have to take into account 
the poles surrounded by the $\lambda_2$-contour 
$-\mathcal{C}\cup\widetilde{\mathcal{C}}$ i.e.
$\lambda_2=-\pi \i/4, z$.
After evaluating the integral with respect to $z$ by 
the method used in the calculation for the
density-density correlation function,
one has
\begin{equation}
\bra c_1 c_{m+1}^{\dagger}\ket
=\frac{-1}{2\pi}\int_{\mathcal{C}}
 \d \lambda \frac{\varphi^m(\lambda)}
                 {\sh(\lambda+\frac{\pi }{4}\i)
                  \sh(\lambda-\frac{\pi }{4}\i)}.
\end{equation}
Changing the variables as explained before, one
finally obtains
\begin{equation}
\bra c_1 c_{m+1}^{\dagger}\ket
=\frac{-1}{2\pi}\int_{k_{\rm F}}^{2\pi-k_{\rm F}}\d k
      \e^{\i m k}
=\frac{\sin(k_{\rm F} m)}{ \pi m}
\qquad k_{\rm F}\in [\frac{\pi}{2},\pi].
\end{equation}
This  also coincides with the well-known result. 
In particular at zero chemical potential $\mu_{\rm c}=0$
($k_{\rm F}=\pi /2$), one sees
\begin{equation}
\bra c_1 c_{m+1}^{\dagger}\ket=
-(-1)^m \frac{\sin(\frac{\pi}{2} m)}{\pi m}.
\end{equation}
%
\section{Summary and discussion}
%
In this paper,  correlation functions for the spinless 
fermion model have been discussed by using the fermionic 
$R$-operator and the algebraic Bethe ansatz.
Applying solutions of the inverse scattering problem, we 
derived determinant representations for  form factors of
local fermion operators.
In addition, multiple integrals representing
the density-density correlation function and
the equal-time one-particle Green's function
were obtained both for arbitrary interaction
strengths and particle densities.
In particular for the free fermion model, these 
formulae reduce to the known exact results.

Before closing this paper, we would like to remark 
possible generalizations of our formulae.
A generalization to the finite temperature
case is of importance.
In fact, the quantum transfer matrix, which 
is a powerful tool for study of finite-temperature
properties for strongly correlated systems in
one-dimension, has already been provided for the 
spinless fermion model \cite{SSSU}.
Because the algebraic relations among the elements
of monodromy operators in the quantum transfer
matrix approach are completely the same as
\eqref{commutation}, the relations derived in
this paper are applicable without essential changes.

It will also be very interesting to compute the
spectral function for the spinless fermion model.
Considering several excited states and inserting the 
corresponding Bethe roots into the determinant 
representations for the form factors \eqref{fp} 
and the scalar product \eqref{sp}, one can
explicitly compute the spectral function 
for a finite system.
The method developed
in the study of the dynamical structure factors
for the XXZ chain (see \cite{BKM02,Sato04,CHM05,CM05,Pereira06,CH06} 
for example) will also be useful in computations of the
spectral function of the spinless fermion model. 
Comparison with the arguments in \cite{KPKG07} is also interesting.
%
\section*{Acknowledgments}
%
The authors would like to thank H. Boos, F. G\"ohmann, A. Kl\"umper 
and M. Shiroishi for fruitful discussions.  This work is partially 
supported by Grants-in-Aid for Young Scientists (B) No.~17740248,
Scientific Research (B) No.~18340112 and (C) No.~18540341 from
the Ministry of Education, Culture, Sports, Science and Technology 
of Japan.
%

\end{document}